\documentclass[11pt]{article} 
 
\usepackage{bm,epsfig,amsmath} 
\usepackage{amssymb}
\usepackage{rotating}

\def\Dirac{\textsc{Dirac}}

\def\ketm#1{  \left\vert  #1   \right\rangle   }  
\def\bram#1{  \left\langle  #1   \right\vert   }  

\def\sprm#1#2{  \left\langle #1 \left\vert \right. #2 \right\rangle   }

\def\procref#1{{\sf #1}}

\def\mem#1#2#3{  \left\langle #1 \left\vert  #2 \right\vert #3 \right\rangle   }

\begin{document} 

%
%

\title{Algebraic tools for dealing with the atomic shell model. \\
       I.\ Wavefunctions and integrals for hydrogen--like ions} 
 
\author{Andrey Surzhykov\footnote{\:To whom correspondence should be 
        addressed (surz@physik.uni-kassel.de)} , 
        Peter Koval and Stephan Fritzsche\footnote{
	\:s.fritzsche@physik.uni-kassel.de}
        \\ 
	\\ 
	\\ 
        Institut f\"ur Physik, Universit\"a{}t Kassel,  \\ 
        Heinrich--Plett--Str. 40, D--34132 Kassel, Germany 
	\\ 
	} 

\date{\today}

\maketitle

\thispagestyle{empty} 
\enlargethispage{0.0cm} 

%
%
 
\begin{abstract} 
Today, the 'hydrogen atom model' is known to play its role not only in teaching 
the basic elements of quantum mechanics but also for building up effective 
theories in atomic and molecular physics, quantum optics, plasma physics,
or even in the design of semiconductor devices. Therefore, the analytical
as well as numerical solutions of the hydrogen--like ions are frequently 
required both, for analyzing experimental data and for carrying out quite 
advanced theoretical studies. In order to support a fast and \textit{consistent} 
access to these (Coulomb--field) solutions, here we present the \Dirac{} 
program which has been developed originally for studying the properties and 
dynamical behaviour of the (hydrogen--like) ions. In the present version, 
a set of \textsc{Maple} procedures is provided for the Coulomb wave and Green's 
functions by applying the (wave) equations from both, the nonrelativistic and
relativistic theory. Apart from the interactive access to these functions, 
moreover, a number of radial integrals 
are also implemented in the \Dirac{} program which may help the user to 
construct transition amplitudes and cross sections as they occur frequently 
in the theory of ion--atom and ion--photon collisions.       
\end{abstract}
%

\newpage 
 
\textbf{\large PROGRAM SUMMARY} 
 
\bigskip

\textit{Title of program:} \Dirac{} 
 
\bigskip

\textit{Catalogue number:} To be assigned. 
 
\bigskip

\textit{Program obtainable from:} CPC Program Library, 
     Queen's University of Belfast, N.~Ireland.  
 
\bigskip

\textit{Licensing provisions:} None. 
 
\bigskip

\textit{Computer for which the program is designed and has been tested:} All
     computers with a license of the computer algebra package 
     \textsc{Maple} [1].
       
\bigskip

\textit{Program language used:} Maple 8 and 9. 
 
\bigskip

\textit{Memory required to execute with typical data:}  
 
\bigskip

\textit{No.\ of bytes in distributed program, including test data, etc.:}  
 
\bigskip

\textit{Distribution format:} tar gzip file
 
\bigskip

\textit{CPC Program Library Subprograms required:} None.
   
\bigskip

\textit{Keywords:} Analytical solution, Coulomb--Green's function, Coulomb
                   problem, Dirac
                   equation, energy level, expectation value, hydrogen--like 
		   ion, hydrogenic wavefunction, matrix element, radial 
		   integral, special functions.   
 
\bigskip

\textit{Nature of the physical problem:}  \newline 
Analytical solutions of the hydrogen atom are widely used in very different
fields of physics [2,3]. Despite of the rather simple structure of the 
hydrogen--like ions, however, the underlying 'mathematics' is not always that
easy to deal with. Apart from the well--known level structure of these ions 
as obtained from either the Schr\"odinger or Dirac equation, namely, 
a great deal of other properties are often needed. These properties are 
related to the interaction of bound electron(s) with external particles 
and fields and, hence, require to evaluate transition amplitudes, including 
wavefunctions and (transition) operators of quite different complexity. 
Although various special functions, such as the Laguerre polynomials, 
spherical harmonics, Whittaker functions, or the hypergeometric 
functions of various kinds can be used in most cases in order to express these 
amplitudes in a concise form, their derivation is time consuming and prone 
for making errors. In addition to their complexity, moreover, there exist 
a large number of mathematical relations among these functions which are
difficult to remember in detail and which have often hampered quantitative 
studies in the past. 

\bigskip

\textit{Method of solution:}  \newline 
A set of \textsc{Maple} procedures is developed which provides both the
nonrelativistic and relativistic (analytical) solutions of the `hydrogen atom
model` and which facilitates the symbolic evaluation of various transition 
amplitudes. 

\bigskip

\textit{Restrictions onto the complexity of the problem:}  \newline 
Over the past decades, a large number of representations have been worked out
for the hydrogenic wave and Green's functions, using different variables and
coordinates [2]. From these, the position--space representation in 
spherical coordinates is certainly of most practical interest and has been used
as the basis of the present implementation. No attempt has been made by us so 
far to provide the wave and Green's functions also in momentum space, for which 
the relativistic momentum functions would have to be constructed numerically.  

Although the  \Dirac{} program supports both symbolic and numerical 
computations, the latter one are based on \textsc{Maple}'s standard 
\textit{software} floating--point algorithms and on the (attempted) precision
as defined by the global {\tt Digits} variable. Although the default number,
{\tt Digits = 10}, appears sufficient for many computations, it often leads to
a rather dramatic loss in the accuracy  of the relativistic wave functions and 
integrals, mainly owing to \textsc{Maple}'s imprecise internal evaluation of 
the corresponding special functions. Therefore, in order to avoid such 
computational difficulties, the {\tt Digits} variable is set to 20 whenever 
the \Dirac{} program is (re--)loaded.

\bigskip

\textit{Unusual features of the program:} \newline 
The \Dirac{} program has been designed for interactive work which, apart from
the standard solutions and integrals of the hydrogen atom, also support
the use of (approximate) \textit{semirelativistic} wave functions for 
both, the bound-- and continuum states of the electron. To provide a fast and
accurate access to a number of radial integrals which arise frequently in
applications, the analytical expressions for these integrals
have been implemented for the one--particle operators 
$r^k$, $e^{-\sigma r}$, $d^m/dr^m$, $j_L(kr)$ as well as for the (so--called)
two--particle Slater integrals which are needed to describe the 
Coulomb repulsion among the electrons. Further procedures of the \Dirac{}
program concern, for instance, the conversion of the physical results between 
different unit systems or for different sets of quantum numbers. A brief description of 
all procedures as available in the present version of the \Dirac{} program is 
given in the user manual \texttt{Dirac-commands.pdf} which is distributed 
together with the code.

\bigskip

\textit{Typical running time:} Although the program replies \textit{promptly}
on most requests, the running time also depends on the particular task.

\bigskip 
 
{\it References:}   \newline 
 [1] Maple is a registered trademark of Waterloo Maple Inc. \newline
 [2] H.~A.~Bethe and E.~E.~Salpeter, Quantum Mechanics of
     One-- and Two--Electron Atoms, Springer, Berlin, 1957. \newline
 [3] J.~Eichler and W.~Meyerhof, Relativistic Atomic Collisions, 
     Academic Press, New York, 1995.
 
          
\newpage 
{\large\bf LONG WRITE--UP} 
 
\bigskip 
 
%
%
\section{Introduction} 

Since the beginning of quantum mechanics, the hydrogen atom has been
one of the best studied models in physics \cite{BeS57,BaJ83}. In a large number
of textbooks and monographs, therefore, its (well--known) solutions from 
either the nonrelativistic Schr\"odinger equation or Dirac's (relativistic)
wave equation have served as standard examples in introducing atomic theory.
But apart from the educational insight as obtained from the
\textit{hydrogen atom model}, the theory of the one--electron atoms and ions 
has found its way also into quite different fields of modern physics 
including molecular and plasma physics, quantum optics, or the theory of 
quantum information. When combined with the \textit{atomic shell model},
i.e.\ the successive filling of shells as established for instance by
the periodic table of elements, the (analytical) solutions of the `hydrogen 
atom` help understand many atomic processes in Nature, at least qualitatively. 
During the last decades, moreover, Dirac's theory of the hydrogen--like ions 
has been applied successfully for describing energetic collisions of high--Z 
ions with atomic targets \cite{EiM95} or for testing quantum--electrodynamical 
(QED) effects \cite{MoP98}, if Furry's picture and field--theoretical concepts 
are taken into account.

\medskip

As seen from this wide--spread use of the hydrogen atom model,
therefore, a fast and interactive access to the solutions and properties of 
the one--electron ions is highly desirable. In the past, several 
program have been developed and implemented into the \textsc{cpc} program 
library for calculating the hydrogenic wave and Green's functions, 
both within the nonrelativistic \cite{NoT84,BeS80} and 
relativistic \cite{SaM92,KoF03} framework. Beside of providing these
functions explicitly, they have been implemented also --- at least to a 
certain extent --- in a large number of further codes including those for
studying the two--photon ionization of hydrogen--like ions \cite{KoF03}, 
their interaction with lasers \cite{MaH99} and electrons \cite{Moo71}, 
the excitation and ionization of such ions by Coulomb fields \cite{LuS01}, 
their bound--bound transitions in the presence of external magnetic fields 
\cite{Kar79}, and at many places elsewhere. But although these programs have 
certainly found useful for understanding the properties and behaviour of
one--electron atoms and ions, they have  two drawbacks in common which 
restrict their further application: Since these programs are designed 
for a very particular task, they often (i) cannot be extended so easily 
to other problems. Moreover, by using standard languages such as 
\textsc{Fortran} or \textsc{C}, these programs 
do (ii) not support any algebraic (symbolic) access to the wave functions 
and matrix element as required by more advanced studies. The lack of symbolic
techniques, in particular, and together with the low flexibility in dealing 
with the input and output of the programs from above have hampered their use 
for a much wider range of applications.

\medskip

Today, a new and promising alternative for having an algebraic
access to complex (symbolic) expressions is offered by computer algebra 
including the general purpose systems such as \textsc{Mathematica} or 
\textsc{Maple}. In the last decade, they have been increasingly utilized
to provide a fast and reliable gateway to a large set of complex expressions 
and computations. With the development of the \Dirac{} program, we here provide 
a set of \textsc{Maple} procedures for studying the properties and the 
dynamical behaviour of the hydrogen--like ions. In the past two years, this 
program has been used to support not only the interpretation of experiments but 
also for a detailed analysis of the polarization and correlation properties 
of the x--ray radiation as emitted by relativistic collisions 
\cite{SuF03, SuF02}. Owing to its interactive design, moreover, the \Dirac{} 
program may help the user not only in his or her daily research work but 
also in teaching the physics of (hydrogen--like) atoms and ions.

\medskip

Obviously, however, a program about the \textit{hydrogen atom model} can 
neither be implemented comprehensively nor explained within a single step. 
Therefore, the \Dirac{} package will be presented in several steps, each of 
them dedicated to a well defined part of the theory. To provide access to
the very basic elements of the theory, here we first describe a set of 
procedures for the symbolic (as well as numerical) computation of the wave 
and Green's functions. These functions are provided for both, the bound-- and 
continuum--state solutions of the nonrelativistic as well as relativistic 
theory. Having once a simple access 
to these functions, of course, they can be rather easily applied to construct 
the transitions amplitudes and matrix elements for a large number of 
other properties. For several --- frequently required --- matrix elements, 
an algebraic evaluation of the radial integrals has been implemented also 
explicitly.

\medskip

Although the main purpose of this work is to describe the design and the
use of the \Dirac{} package, in the next section we first compile the basic
equations and expressions from the theory of the hydrogen atom with emphasis on
those which have been implemented in the code. In Section \ref{s:lib-descr},
we later describe the organization of the program and how it is distributed.
This sections also lists all the user--accessible commands together with a
short description, while a detailed explanation of the parameters, optional
arguments and some additional information for each procedure is given in an
additional user manual and is appended to the code. This is followed in Section
\ref{examples} by a number of selected examples to illustrate the interactive
use of the \Dirac{} program within its \textsc{Maple} environment. A short
outlook onto future developments and extensions of the program is finally given
in Section \ref{outlook}. 

%
%
%
%
%
%
%
\section{Theoretical background}

Since the quantum--mechanical theory of the one--electron atoms has been worked
out long time ago, both within the nonrelativistic and relativistic framework,
we may refer for most details to the literature. Helpful introductions into this
topic can be found, for instance, in the texts by Bethe and Salpeter 
\cite{BeS57}, Messiah \cite{Mes99}, Bransden and Joachain \cite{BaJ83}, 
Eichler and Meyerhof \cite{EiM95}, and by many others. In the present
work, therefore, we will restrict ourselves to rather a short compilation of 
the basic expressions, just enough in order to introduce the notations and 
formulas as implemented within the \Dirac{} program. In the following two
subsections \ref{theory_wave_functions} and \ref{theory_Green_functions}, 
in particular, we briefly recall the (analytically well known) radial--angular
representation of the hydrogen wave and Green's functions which are later 
applied in subsection \ref{theory_integrals} for evaluating a number of radial 
integrals. Emphasis has first been paid to those integrals which are of frequent
use in the computation of the structural and dynamical properties of the 
hydrogen--like ions. In this paper (and also in most parts of the underlying 
implementation of the \Dirac{} code), atomic Hartree units 
$\hbar$ = $m_e$ = $e^2 / 4 \pi \epsilon_0$ = 1, where $\epsilon_0$ is the
permittivity of free space, are used throughout unless specified otherwise.

\subsection{Wavefunctions and energy levels}
\label{theory_wave_functions}

\subsubsection{Nonrelativistic framework} 

In a pure Coulomb potential $V(r) = - Z/r$ of a given (nucleus with)
charge $Z$, the motion of an electron is described by the time--independent 
Schr\"o{}dinger equation 
\begin{eqnarray}
   \label{nonrelativistic-Schroedinger}
   \Delta \psi(\mathbf{r}) + 
   2 \left( E + \frac{Z}{r} \right) \psi(\mathbf{r}) & = & 0 \, 
\end{eqnarray}
which, in polar coordinates $ r, \, \vartheta,$ and $\varphi $, can be 
separated into three independent equations owing to the ansatz \cite{BeS57}
\begin{eqnarray}
   \label{non_rel_function}
   \psi (r, \vartheta, \varphi ) & = & R(r) \, Y_{lm} (\vartheta, \varphi ) 
   \; = \; \frac{P(r)}{r} \, Y_{lm} (\vartheta, \varphi ) \, .
\end{eqnarray} 
As seen from this ansatz (\ref{non_rel_function}), the electronic wavefunction 
$\psi(\mathbf{r}) \equiv \psi(r, \theta, \phi) $ is therefore given as a 
product of a \textit{radial} function $\,R(r)\,$ or $\,P(r)/r$, respectively,
and a spherical harmonic $Y_{lm} (\vartheta, \varphi )$ which describes the 
\textit{angular} structure and which depends on the orbital angular momentum 
$l$ and its projection onto the quantization axis as denoted by the magnetic 
quantum number $m$. However, not much need to be said here about the 
properties and treatment of the spherical harmonics as one of the standard 
tasks from the theory of angular momentum \cite{VaM88} for which meanwhile 
also a number of symbolic tools are available \cite{InF01,Fri:03}. 
The (radial) function $P(r)$ from ansatz (\ref{non_rel_function}) obeys the 
\textit{radial} Schr\"o{}dinger equation
\begin{equation}
   \label{radial_Schroedinger}
   \left[ \frac{1}{r^2} \frac{\partial}{\partial r} 
   \left( r^2 \frac{\partial}{\partial r} \right) -
   \frac{l (l + 1)}{r^2} + \frac{2Z}{r} + 2E \right] \frac{P(r)}{r} = 0 \, 
\end{equation}
and are known to belong either to the discrete part of the spectrum with the
(negative) energies 
\begin{eqnarray}
   \label{non_rel_energy}
   E_n & = & - \frac{Z^{\,2}}{2\,n^2} \;<\; 0, \hspace*{1.5cm}
   n \,=\, 1, 2, ... \, ,
\end{eqnarray}
the so--called \textit{bound} states, or to its continuous part for all 
energies $E >$ 0, for which the functions $P(r)$ are said to represent 
the \textit{free--electron} (or \textit{continuum}) states. For a 
Coulomb potential in Eq.\ (\ref{radial_Schroedinger}), an analytic solution of
the radial functions $P(r)$ in terms of the confluent hypergeometric function 
$F(\alpha, \beta; x)$ are known for both, the bound--states 
\begin{eqnarray}
   \label{non_rel_wave_function_bound}
   P_{nl} (r)  & = & r^{l+1} \,
   \frac{1}{(2l+1)!} \, \sqrt{\frac{(n+l)!}{(n-l-1)! \, 2n}} \,
   \left( \frac{2Z}{n} \right)^{3/2+l} \, 
   e^{\left( - \frac{Zr}{n} \right)} \,
   \nonumber  \\[0.2cm]
   &  & \hspace*{4.5cm} \times \:
   F \left( - (n-l-1), \, 2l+2, \, \frac{2Zr}{n} \right) \, , 
\end{eqnarray}

with $n$ and $l$ being the principal and the orbital angular momentum quantum
numbers, as well as for the \textit{continuum} ($E > 0$)
\begin{eqnarray}
   \label{non_rel_wave_function_free}
   P_{\,E l} (r) & = & \frac{2 \sqrt{Z}}{\sqrt{1 - e^{-2 \pi n'}}} \;
   \prod_{s=1}^{l} \, \sqrt{ s^2 + n^{'2}} \,
   \frac{(2kr)^{\,l}}{(2l+1)!} \, e^{-ikr} \;
   F ( i n' +l+1, 2l+2,2ikr ) \, 
\end{eqnarray}
with $ k \: = \: \sqrt{2 E} $ and $ n' \: = \: Z/k $. In the standard theory,
moreover, the radial functions (\ref{non_rel_wave_function_bound}) and
(\ref{non_rel_wave_function_free}) are normalized due to the conditions
\begin{equation}
   \label{non_rel_wave_function_bound_normalization}
   \int_0^{\infty} R^{\,2}_{\,nl}(r) \, r^2 \,dr \;=\; 
   \int_0^{\infty} P^{\,2}_{\,nl}(r) \, dr \;=\; 1
\end{equation}
for having a single particle per bound state and due to
\begin{equation}
   \label{non_rel_wave_function_free_normalization}
   \int_0^{\infty} P^*_{El}(r) \, P_{E`l}(r) \, dr \;=\; \delta(E - E') \,
\end{equation}
for representing one particle per unit energy as far as electrons in the 
continuum are concerned.

\medskip

In the \Dirac{} package, both the full wavefunctions (\ref{non_rel_function}) 
and their radial components (\ref{non_rel_wave_function_bound}) and 
(\ref{non_rel_wave_function_free}) are provided for the  bound-- and 
free--electron states by using the procedures \procref{Dirac\_orbital()}
and \procref{Dirac\_radial\_orbital()}, respectively.

\subsubsection{Relativistic framework}
\label{wave_functions_relativistic}

Similar to Schr\"o{}dinger's equation (\ref{nonrelativistic-Schroedinger}), 
the Dirac (eigenvalue) equation with a Coulomb potential 
\begin{eqnarray}
   \label{rel_Dirac_equation}
   \left(  i c \bm{\alpha} \cdot \bm{\nabla} \, + \, \frac{Z}{r} \, - \,
   \, \left( \beta - 1 \right) c^2 + E \right) \psi (\mathbf{r}) \; = \; 0 \, 
\end{eqnarray}
can be used to describe the \textit{relativistic} motion of a spin--$1/2$ 
electron in the field of a nucleus with charge $Z$ which, apart from the 
kinetic and potential energy of the electron, now also contains its rest energy 
as well as contributions due to the electron spin \cite{BaJ83, EiM95}. 
Analogue to the nonrelativistic case, the solution $\psi(\mathbf{r})$ of 
Eq.\ (\ref{rel_Dirac_equation}) can be separated again into a radial and 
angular part
\begin{eqnarray}
   \label{rel_wave_function_spinor}
   \psi_{\kappa m} (r,\vartheta,\varphi) & = & 
   \frac{1}{r} \, \left( \begin{array}{r} 
         P(r) \, \Omega_{\kappa m}   (\vartheta,\varphi)  \\[0.3cm]
    i \, Q(r) \, \Omega_{-\kappa m}  (\vartheta,\varphi)
                         \end{array} \right) \, ,
\end{eqnarray}
where $ \Omega_{\kappa m} (\vartheta,\varphi) $ denotes a standard Dirac
spin--orbital function and where $\kappa \:=\: \pm$ ($j + 1/2$) for 
$l \:=\: j \pm 1/2$ is called the \textit{relativistic} angular momentum quantum
number. Owing to the definition of $\kappa \,=\, \pm 1, \, \pm 2,\, ...$, 
this quantum number carries information about both, the total angular 
momentum $j$ and the parity (--1)$^l$ of the wavefunction. As usual, 
the Dirac spin--orbitals can be written in terms of the spin--1/2 Pauli 
spinors $ \chi_{\pm} $ and the spherical harmonics by
\begin{eqnarray}
  \label{rel_spin_orbitals}
  \Omega_{\kappa m}(\vartheta,\varphi)  & = & \sum_{m_l} \,
  \sprm{l, m_l, 1/2, m-m_l}{jm} \: Y_{l m_l} (\vartheta,\varphi) \, 
  \chi_{m - m_l} \, .
\end{eqnarray} 
As seen from ansatz (\ref{rel_wave_function_spinor}), moreover, the radial 
part of the wavefunctions is now given by the two functions $P(r)$ and $Q(r)$ 
which are called the \textit{large} and \textit{small} components, respectively.
Similar to the nonrelativistic case, an analytic representation of these 
two components is known and is different for an electron in a bound or 
continuum state where, apart from the continuum with \textit{positive} energies
$E \,>\, 0$  
an additional continuous branch (of solutions) with \textit{negative} energies 
$E \le -2c^2$ exists. As known from Dirac's theory, these solutions with
negative energies require special care and can be re--interpreted, for instance,
in terms of positron states. For the \textit{bound} states of hydrogen--like 
ions, the (discrete eigen--) energies of Eq.\  (\ref{rel_Dirac_equation}) 
are given by
\begin{eqnarray}
   \label{rel_energy}
   E_{n \kappa} & = & \frac{\left( W_{n \kappa} - 1 \right)}{\alpha^{2}} =
   \frac{1}{ \alpha^2 
             \left[ 1 \, + \,\left( \frac{\alpha Z}{n' \,+\, s} \right)^2 
   \right]^{1/2}} - \frac{1}{\alpha^2} \, ,
\end{eqnarray}
while the corresponding radial components $P_{n \kappa}$ and $Q_{n \kappa}$
can be written again in terms of the confluent hypergeometric function 
$ F (\alpha, \beta ; x) $. They read as 
\begin{eqnarray}
\label{hydrogenic-Pnkappa}
   P_{n \kappa} (r) & = & N_{n \kappa} \, \sqrt{1 + W_{n \kappa}} \, \, r \, 
   (2qr)^{s-1} \, e^{-qr} \, 
   \left[ -n' \, F (-n'+1, 2s+1 ; 2qr) \; - \;  
   \right.  \nonumber  \\[0.3cm]  &  & \hspace*{5.3cm} \left. 
          \left( \kappa \, - \, \frac{\alpha Z}{q \lambda_c} \right) \, 
          F (-n', 2s+1 ; 2qr) \right]  \, ,
   \\[0.8cm]
\label{hydrogenic-Qnkappa}
   Q_{n \kappa} (r) & = & -  N_{n \kappa} \, \sqrt{1 - W_{n \kappa}} \, \, r \, 
   (2qr)^{s-1} \, e^{-qr} \, 
   \left[ n' \, F (-n'+1, 2s+1 ; 2qr) \; - \; 
   \right.  \nonumber  \\[0.3cm]  &  & \hspace*{5.3cm} \left.
          \left( \kappa \, - \, \frac{\alpha Z}{q \lambda_c} \right) \, 
          F (-n', 2s+1 ; 2qr) \right]  \, .
\end{eqnarray}
where $n' \, = \, n - | \kappa | \, = \, 0, \, 1, \, 2, \, ... \,$ denotes the
number of nodes of the radial components, $ \lambda_c \,=\, \hbar/ m_e c$ 
the Compton length of the electron, and
\begin{eqnarray}
   \label{q_s_coefficients}
   s & = & \sqrt{\kappa^2 \, - \, (\alpha Z)^2} \, , \\
   q & = & \frac{Z}{ \sqrt{(\alpha Z)^2 \, + \, (n' + s)^2} } \, .
\end{eqnarray}
Moreover, the normalization factor 
\begin{eqnarray}
   \label{N_normalization}
   N_{n \kappa} & = & 
      \frac{\sqrt{2} \,q^{\,5/2} \,\lambda_c}{\Gamma (2s+1)} \;
      \left[ \frac{\Gamma (2s+n'+1) }{
                   n'! \: (\alpha Z) \, 
                   (\alpha Z \, - \, \kappa q \lambda_c)} \right]^{1/2} \, 
\end{eqnarray}
is chosen in such a way that the two radial components satisfy the condition
\begin{equation}
   \label{rel_wave_function_bound_normalization}
   \int_0^{\infty} \left( P^{\,2}_{n \kappa}(r) \,+\, 
                          Q^{\,2}_{n \kappa}(r) \right) \,
   dr \: = \: 1 \, ,
\end{equation}
i.e.\ that the wavefunction (\ref{rel_wave_function_spinor}) represents one 
electron per bound state.

\medskip

For the free--electron states with energies $E \le -2c^2\,$ or $\,E >$ 0,
the explicit expressions of the radial components $P_{E \kappa}(r)$ and 
$Q_{E \kappa}(r)$ are quite elaborate when compared with the bound solutions
in Eqs.\  (\ref{hydrogenic-Pnkappa}) and (\ref{hydrogenic-Qnkappa}) and are
not displayed here. They are shown and discussed, for example, in the text 
by Eichler and Meyerhof \cite{EiM95}. However, these continuum states are also
implemented in the \Dirac{} package and can be accessed as well by means of 
the procedures \procref{Dirac\_spinor()} and 
\procref{Dirac\_radial\_component()}.

\subsubsection{Semirelativistic framework}

The two radial functions (\ref{hydrogenic-Pnkappa}) and 
(\ref{hydrogenic-Qnkappa}) represent, together with Eq.\ 
(\ref{rel_wave_function_spinor}), an \textit{exact} solution of the Dirac 
equation for a pure Coulomb potential. In practice, however, the use of these
radial components often leads to rather lengthy and complicated computations. 
Instead of the exact form, therefore, the \textit{approximate} Darwin 
wavefunctions \cite{Dar28} have been widely used in applications such as the 
theoretical description of the Coulomb ionization \cite{Voi04} 
or the electron--positron pair creation \cite{Bec87}. For a Coulomb potential,
$ -Z/r$, the Darwin wavefunctions 
\begin{eqnarray}
   \label{semi_rel_wave_function_Darwin}
   \phi^{(\pm)}_{nlm} (\mathbf{r}) & = &
   \left( 1 \, - \, \frac{i}{2c} \, \bm{\alpha} \cdot \bm{\nabla} 
   \right) \,
   u^{\, (\pm)} \, \psi_{nlm} (\mathbf{r})
\end{eqnarray}
are expressed in terms of the nonrelativistic solutions 
(\ref{non_rel_function}--\ref{non_rel_wave_function_bound}) of the 
Schr\"o{}dinger equation from subsection 2.1.1, where 
$ u^{(+)} \, = \, (1,0,0,0)^+ $ and $ u^{(-)} \, = \, (0,1,0,0)^+ $
are the two four--component spinors for a particle at rest with either spin--up 
or spin--down, respectively. Obviously, the Darwin functions 
(\ref{semi_rel_wave_function_Darwin}) have a much simpler structure when
compared with the exact bound--state solutions (\ref{hydrogenic-Pnkappa}) and
(\ref{hydrogenic-Qnkappa}) from above; they are accurate to first--order in 
$\alpha Z$ in the expansion of the Dirac
equation and are normalized to the same order.

\medskip

Apart from the bound states (\ref{semi_rel_wave_function_Darwin}),
approximate continuum wavefunctions can be derived which are accurate to 
$\alpha Z$ in the relativistic motion of the electron and which are
known as Sommerfeld--Maue wavefunctions \cite{SoM35} from the literature.
Again, these functions can be expressed in terms of the nonrelativistic functions
and have been widely used in applications \cite{Bec87, BeM54}. Although the
explicit form of these functions is slightly more complicated than in the bound
case, their derivation follows very similar lines. Here, we will not display 
these functions in detail but simply refer to the procedures 
\procref{Dirac\_Darwin()} and \procref{Dirac\_Sommerfeld\_Maue()} of the 
\Dirac{} code which facilitate 
the access to both, the Darwin and the Sommerfeld--Maue semirelativistic 
approximations.

\subsection{Green's function}
\label{theory_Green_functions}

Apart from the wavefunctions, which describe the electron in some 
particular quantum state, one often needs the summation over all the 
(unoccupied) states, especially if parts of the atomic interaction are treated 
as a perturbation. 
Such a summation over the \textit{complete} spectrum is required, for example, 
for describing the two-- (and multi--) photon processes in hydrogen--like 
ions \cite{KoF04} or for the computation of hyperpolarizabilities. Although, 
in principle, it appears straightforward to carry out such a summation 
explicitly by using the function from above, the large number of terms and 
the need for integrating over the electron continuum may hamper
such an approach. Instead, the use of the Green's function
\begin{equation}
   \label{non_rel_Green_function}
   G({\mathbf r}, {\mathbf r'}; E) \ = \ \sum\mkern-22mu\int_{\nu} \  
   \frac{\ketm{\psi_{\nu}({\mathbf r})} \bram{\psi_{\nu}({\mathbf r'})}}
   {E_\nu - E} \, 
\end{equation}
is usually much more favorable since it provides a simple access to the
perturbative treatment of various types of interactions in the computation of
atomic properties. 

\medskip

As easily shown, the Green's function (\ref{non_rel_Green_function}) can be
obtained generally as solution of the inhomogeneous differential equation 
\begin{eqnarray}
   \label{non_rel_Greens_equation}
   \left[ \hat{H} - E \right] \:
   G(\mathbf{r},\mathbf{r'}; E) & = & 
   \delta (\mathbf{r} - \mathbf{r'}) \; ,
\end{eqnarray}
where $\hat{H}$ denotes the Hamiltonian operator of the system. Similar to the
case of the wavefunctions, the form of the Hamiltonian $\hat{H}$ and, hence, the 
explicit expressions for the Green's function depends of course on the 
\textit{framework} in which the system is described. For the Schr\"odinger
operator $\,\hat{H} \,=\, - \nabla^2/2 \:-\: V(r)\,$ (the nonrelativistic case), 
for example, the  solutions of Eq.\ (\ref{non_rel_Greens_equation}) can be 
separated  -- again -- into radial and angular functions by means of the 
ansatz
\begin{eqnarray}
   \label{non_rel_Greens_function}
   G(\mathbf{r},\mathbf{r'}; E) & = & 
   \frac{1}{r r'} \: \sum_{l = 0}^{\infty} \: \sum_{m} \:
   g_l (r,r'; E) Y_{lm}^* \: (\vartheta', \varphi') \:
   Y_{lm} (\vartheta, \varphi) \, .
\end{eqnarray}
For a Coulomb--potential $V(r) \,=\, -Z/r$ in the Hamiltonian above, 
the radial functions are known analytically in terms of (special) Whittaker 
functions of the first and second kind

\begin{eqnarray}
   \label{non_rel_Greens_function_radial}
   g_l (r,r'; E) & = & \frac{\nu}{Z}\: 
   \frac{\Gamma (l+1-\nu)}{\Gamma (2l+2)} \, 
   M_{\nu,l+1/2} \left( \frac{2 r_< Z}{\nu} \right)  \;
   W_{\nu,l+1/2} \left( \frac{2 r_> Z}{\nu} \right)  ,
\end{eqnarray}
where $ \nu \, = \, \frac{Z}{\sqrt{-2 E}} $ and $r_>$ = max($r$, $r'$) and where
$r_<$ = min($r$, $r'$) refer to the larger and smaller values of the two radial
coordinates, respectively.  

\medskip

A similar \textit{radial--angular} representation can be found for the
(relativistic) Dirac Hamiltonian in Eq.\ (\ref{non_rel_Greens_equation}) 
but leads to a quite complicated form for the radial part which is given 
then by a $2 \times 2$ matrix. In the \Dirac{} program, we provide the 
procedure \procref{Dirac\_Greens\_{}radial()} which supports the computation of
the radial functions (\ref{non_rel_Greens_function_radial}) or 
of the corresponding relativistic Green's functions.

\subsection{Integrals with hydrogenic wavefunctions}
\label{theory_integrals}

The Coulomb wave and Green's functions from the previous subsections can be
applied for studying the properties and dynamics of hydrogen--like ions.
Usually, they are the basic ingredients in defining the transition amplitudes 
and cross sections and, by using their \textit{radial--integral} decompositions
in Eqs.\  (\ref{non_rel_function}), (\ref{rel_wave_function_spinor}), and 
(\ref{non_rel_Greens_function}) immediately lead to a number of 
\textit{radial integrals} including various types of (transition) operators.
While, of course, these integrals can always be determined \textit{numerically}, 
an \textit{analytic} (pre--) evaluation of these integrals is in many cases
possible and results in a considerable gain in the computation of the transition
amplitudes. Often, moreover, the analytic evaluation of the radial integrals
is faster and more accurate when compared with a (straightforward) numerical 
integration. For several integrals, which frequently arise in applications,
therefore, the analytic formulas have been implemented explicitly into the
\Dirac{} code. Below, we first discuss the matrix elements containing the 
operators $r^k$, $e^{-\sigma r}$, $d^m/dr^m$, the integrals for studying the
light--matter interaction or for collisions with external particles, but turn
later also to the (well--known) Slater integrals for treating the interaction 
among the electrons.

\subsubsection{Radial integrals of $r$--dependent operators}

\textit{Nonrelativistic framework:} For calculating the structure of the
hydrogen--like ions and their cross sections in collisions with atoms and
electrons, one often needs the matrix elements for the operators 
$r^k$, $d^m/dr^m$, $e^{-\sigma r}$, or various combinations of these operators.
Within the nonrelativistic Schr\"odinger theory, the analytic calculation of 
such $r$--dependent matrix elements have been analyzed and discussed 
especially by S\'anchez and L\'opez Pin\~eiro \cite{SaP93} who
applied the hypervirial theorem along with a second quantization formalism.
They showed that these matrix elements can be typically reduced to radial
integrals of the form

\begin{small}
\begin{eqnarray}
   \label{r-matrix-initial}
   \mem{n \, l}{r^k \, e^{-\sigma r}}{n' \, l'} 
   & = & 
   \int\limits_{0}^{\infty}{P_{n l}(r) \, r^k \, e^{-\sigma r} \, P_{n' l'}(r) \, dr}
   \nonumber \\[0.25cm] \hspace{-3.0cm} 
   &  & \hspace{-3.0cm} =\;
   N_{n' l'} \, N_{n l} \, \sum\limits_{m = 0}^{n - l - 1} \, 
   \sum\limits_{m' = 0}^{n' - l' - 1} \, \frac{(-1)^{m + m'}}{m! \, m'!} 
    \binom{n' + l'}{n' - l' -1 - m'}  \, 
   \binom{n + l}{n - l -1 - m}  \, 
   \nonumber \\[0.25cm]
   &\times &
   \left( \frac{2Z}{n'} \right)^{l' + m'} \, 
   \left( \frac{2Z}{n} \right)^{l + m} 
   \left( \frac{Z}{n'} + \frac{Z}{n} + \sigma \right)^{-(k + l + l' + m + m' +3)}
   \nonumber \\[0.25cm]
   &\times &  \, \Gamma(k + l + l' + m + m' + 3) \, ,
   \\[-1.0cm] \nonumber
\end{eqnarray}
\end{small}

which can then be utilized also (if necessary) in order to compute the matrix 
elements including some derivatives $d^m/dr^m \; P_{nl} (r)$ of the wave 
functions. For the sake of brevity, however, we will not display these 
recursion formulas but only mention that, for all the operators from above, 
the corresponding radial integrals can be evaluated by means of the procedure 
\procref{Dirac\_r\_matrix\_element()} of the \Dirac{} program. In this
procedure, an analytic evaluation of the integrals is carried out whenever 
possible.
 
\bigskip

\textit{Relativistic framework:}
Of course, an analytical expression of the matrix elements with operators 
$r^k \, e^{-\sigma r}$ can be found also for the relativistic Dirac wavefunctions.
However, in order to derive such expressions it is more convenient not 
to start from the standard representation of the hydrogenic functions 
(\ref{hydrogenic-Pnkappa}--\ref{hydrogenic-Qnkappa}) but first to re--write 
them in terms of a series expansion in $r$ as it was suggested originally
by Rose \cite{Ros61}. By using such $r$--series representation, the 
$r^k \, e^{-\sigma r}$ matrix elements can be written, similar to 
(\ref{r-matrix-initial}), in term of $\Gamma$-- functions as 
\begin{eqnarray}
   \label{r-matrix-initial-relativistic}
   \mem{n \, \kappa}{r^k \, e^{-\sigma r}}{n' \, \kappa'} & = & 
   \int\limits_{0}^{\infty}{ \left( P_{n l}(r) \, P_{n' \kappa'}(r) +
   Q_{n l}(r) \, Q_{n' \kappa'}(r) \right) \, r^k \, e^{-\sigma r} \, dr} \nonumber \\
   & = & N_{n \kappa} \, N_{n' \kappa'} \, \sum\limits_{\nu = 0}^{n - |\kappa|}
   \, \sum\limits_{\nu' = 0}^{n' - |\kappa'|} \, \left( c^+_{n \kappa, \nu} 
   c^+_{n' \kappa', \nu'} +
   c^-_{n \kappa, \nu} c^-_{n' \kappa', \nu'} \right) 
   \nonumber \\[0.25cm] 
   &   & \hspace*{0.0cm} \times \; \frac{q^{s + \nu - 1} \, 
   q'^{s' +\nu' - 1} }{(q + q' + \sigma)^{1 + \nu + \nu' + s + s' + k}} \, 
   \Gamma( \nu + \nu' + k + s + s' + 1) \, ,
\end{eqnarray}
where the parameters $q,\; s$ and the normalization factor $N_{n \kappa}$ were
defined by Rose [26], and where
the coefficients $c^{\,\pm}_{n \kappa, \nu}$ are given by
\begin{equation}
   c^{\,\pm}_{n \kappa, \nu} \; = \; 
   \left( 1 \pm \sqrt{1 - q^2} \right)^{1/2} \,
   \frac{((-n + |\kappa|)_\nu \, 2^\nu)}{\nu! \: (2s + 1)_\nu} \,
   \left[ (\nu - n + |\kappa|) \pm (\alpha Z/q - \kappa ) \right] \, 
\end{equation}
with $\,(a)_\nu\,$ being the Pochhammer symbol.

\medskip

In contrast to the radial integral (\ref{r-matrix-initial-relativistic}), no 
analytic expressions for the matrix elements containing the derivative 
$d^m/dr^m$ within the relativistic framework are implemented explicitly in 
\Dirac{} program. For all such matrix elements, however, the (straightforward) 
numerical integration is applied by making use of the 
standard \textsc{Maple} command {\tt int()}.
 
\newpage
\subsubsection{Grant's integrals}

\textit{Relativistic framework:} As known from relativistic collision theory 
\cite{EiM95, Gra74}, the interaction of (atomic) electrons with the radiation
field or with other atoms and ions can usually be traced back to the computation
of the following three radial integrals 
\begin{eqnarray}
\label{ILplusminus}
   I_L^{\,\pm} (q;a,b) 
   & = & 
   \int_0^\infty dr \, j_L (qr) \;
   \left(  P_a (r) Q_b(r) \, \pm \, Q_a (r) P_b(r) \right) \,
   \\[0.25cm]
\label{IL0}
   I_L^{\,0} (q;a,b) 
   & = & 
   \int_0^\infty dr \, j_L (qr) \; P_a (r) Q_b(r) \,
   \\[0.25cm]
\label{JL}
   J_L  (q;a,b) & = &
   \int_0^\infty dr \, j_L (qr) \;
   \left(  P_a (r) P_b(r) \, + \, Q_a (r) Q_b(r) \right) \, 
\end{eqnarray}
where $j_L(qr)$ denotes a spherical Bessel function of order $L$ and,
as before, $P (r)$ and $Q (r)$ are the radial components of the 
one--electron wavefunctions (\ref{rel_wave_function_spinor}). 
Because of the oscillations in the Bessel as well as the radial wavefunctions, 
however, an accurate value of these integrals is not 
always easy to obtain numerically. In the \Dirac{} program, we therefore
implemented the analytic formulas for these integrals in terms of a finite
series over the confluent hypergeometric functions as known for the 
\textit{bound--bound} and \textit{bound--free} transitions \cite{TrB83, VaB84}.
While this typically results in a much faster (and more accurate) computation 
of the integrals, it requires however that at least one set of radial 
components --- either $P_{a}(r),\; Q_{a}(r)$ or $P_{b} (r),\; Q_{b} (r)$ --- 
has to represent some bound--state electron.

\medskip

\textit{Nonrelativistic framework:} Since, in the limit 
$c \rightarrow \infty$, 
the small component $Q (r)$ becomes negligible, only contributions from 
the last integral (\ref{JL}) survives in the nonrelativistic description of 
interaction processes with the radiation field and gives rise to the integral 
\begin{equation}
   \label{non_rel_Grant_integrals}
   J_L  (q;a,b) =  
   \int_0^\infty dr \, j_L (qr) \: P_a (r) P_b(r) \, .
\end{equation}
An analytic evaluation of this integral with nonrelativistic radial orbital
functions is supported by the procedure \procref{Dirac\_IJ\_radial()} 
of the \Dirac{} program for both, \textit{bound--bound} and 
\textit{bound--free} electronic transitions.

\subsubsection{Slater integrals}

Although the main emphasis in developing the \Dirac{} program has been placed
on studying the properties and behaviors of the hydrogen--like ions, the wave
functions from above can be used also to estimate the effects of the
electron--electron interaction for \textit{few--electron} atoms and ions. 
For such systems, one often needs to compute the (two--electron) matrix 
elements for the Coulomb repulsion operator $1/r_{12}$ which, after some 
algebra, leads to the so--called Slater integrals 
\begin{equation}
   \label{non_rel_Slater_integrals}
   R^{\,k} (a, b, c, d) = \int_0^\infty dr_1 \int_0^\infty dr_2 \;
   P_{n_a l_a}(r_1) \: P_{n_b l_b}(r_2) \:\frac{r_<^{\,k}}{r_>^{\,k+1}} \:
   P_{n_c l_c}(r_1) \: P_{n_d l_b}(r_2) \, ,
\end{equation}
where $r_> \,=\, \max (r_1, r_2)$ and $r_< \,=\, \min (r_1, r_2)$ refer to 
the larger and smaller values of the two radial coordinates, respectively. 
The analytic evaluation of this integral in terms of confluent hypergeometric
functions has been implemented in the \Dirac{} package for
the nonrelativistic framework. For reasons of efficiency, however, no attempt 
has been made to incorporate also the 'relativistic version' of the Slater 
integrals which would raise much larger demands regarding the accurate 
computation of these values.

\section{Program organization} 
\label{s:lib-descr}

\subsection{Overview about the D{\small IRAC} package} 
\label{organization_general}

The \Dirac{} package has been developed as an interactive tool within the
framework of \textsc{Maple} (version 8). It is designed in order to provide 
a simple and consistent access to the wave functions, the excitation and decay
properties as well as the dynamical behaviour of the hydrogen--like ions. 
To allow for a rather large flexibility in the use of the program, most 
of the functions and quantities from the \textit{hydrogen atom model} 
are supported both, within the nonrelativistic 
as well as the relativistic theory. With the present version of \Dirac{}, 
we start with compiling different (analytical) representations of the
Coulomb wave and Green's functions using polar coordinates. In addition, 
both the analytical and numerical evaluation of the -- most frequently -- 
required integrals are supported and now facilitate their accurate computation 
for a large number of properties.

\medskip

Following \textsc{Maple}'s philosophy, the \Dirac{} program is organized 
in a \textit{hierarchical} order. In the current version, there are about 50 
procedures at quite different level of complexity from which most however are
kept hidden to the user. As in \textsc{Maple}'s own design, these procedures 
mainly serve as language elements in order to build up commands at some 
higher level of the hierarchy. 
Therefore, only about 15 procedures at top level need to be known
by the user and are briefly summarized in Table 1. 
More detailed information about the 
use of these procedures, their optional arguments, or the output can be
obtained from the user manual \texttt{Dirac-commands.pdf} which is distributed 
together with code. In this manual, we mainly follow the style of the 
help pages of \textsc{Maple} and \textit{The Maple Handbook} by
Redfern \cite{Red96} from earlier years. To illustrate some of the basic 
features of the program, moreover, a few selected examples are displayed below.

\medskip

As known from \textsc{Maple}'s recent up--grades, most of its internal 
commands make use of rather short names and often of some abbreviations for the
various quantities and methods. Although such a name convention might be 
favorable for the frequent use of the commands, these names are usually
not easy to remember and to recognize in their output if implemented in 
a code. In the \Dirac{} program, therefore, we follow a slightly different 
concept by making use of rather long names to better explain the purpose
of the procedures. In the long run, hopefully, this will
simplify the readability and the maintenance of the program. Moreover, all 
the commands of the \Dirac{} package begin with the additional prefix 
\texttt{Dirac\_} in order to distinguish them from \textsc{Maple}'s
internal functionality.

\begin{table}
\begin{small}
\textbf{Table 1:} Main commands of the \Dirac{} package. A more detailed 
description of these and a few less important procedures is given in
the user manual \texttt{Dirac-commands.pdf} and is distributed together 
with the code. 

\begin{center}
\begin{tabular}{p{4.1cm} p{11.0cm}} \\[-0.4cm]    
\hline \hline  \\[-0.4cm]
Procedure                 & Description 
\\[0.1cm]  \hline  \\[-0.2cm]
\procref{Dirac\_convert()}
   & Converts the physical results between different unit systems. \\[0.1cm]
   
\procref{Dirac\_energy()}       
   & Returns the nonrelativistic energy $E_n$ from Eq.\ (\ref{non_rel_energy}) 
     or the relativistic energy $E_{n \kappa}$ from Eq.\ (\ref{rel_energy})
     for bound--electron states.  \\[0.1cm]

\procref{Dirac\_evalf()}        
   & Attempts to simplify an algebraic expression for either the energies,
   wavefunctions, or matrix elements by substituting the numerical values
   for some fundamental constants. \\[0.1cm]
     
\procref{Dirac\_Greens\_{}radial()} 
   & Calculates the radial part of either the nonrelativistic Coulomb Green's
     functions (\ref{non_rel_Greens_function_radial}) or of the corresponding
     relativistic Green's functions. \\[0.1cm]

\procref{Dirac\_IJ\_radial()}   
   & Evaluates the radial integrals 
     (\ref{ILplusminus}--\ref{non_rel_Grant_integrals}) for both, the 
     bound--bound and bound--free electron transitions. \\[0.1cm]
     
\procref{Dirac\_orbital()}      
   & Returns the nonrelativistic hydrogenic orbitals 
     $\psi_{n (E) \, l \, m} (r, \vartheta, \varphi )$ from Eq.\ 
     (\ref{non_rel_function}) as function of the given arguments. \\[0.1cm]
   
\procref{Dirac\_r\_matrix\_element()}  
   & Evaluates the radial matrix elements for various r--dependent operators 
     between hydrogenic wavefunctions, both within the nonrelativistic and
     the relativistic framework. \\[0.1cm] 
     
\procref{Dirac\_radial\_component()} 
   & Returns the large and small radial components, $P_{n (E) \, \kappa}(r)$ 
     and \newline
     $Q_{n (E) \, \kappa}(r)$, of the relativistic orbital functions
     (\ref{hydrogenic-Pnkappa}--\ref{hydrogenic-Qnkappa}). These components are
     supported for both, the bound and free--electron states. \\[0.1cm] 
    
\procref{Dirac\_radial\_orbital()} 
   & Returns a nonrelativistic radial orbital $P_{n (E) \, l \, m}(r)$ for 
     both, the bound and free--electron states. \\[0.1cm] 

\procref{Dirac\_settings()}        
   & Specifies the framework, nuclear charge and the  units which are to be 
     used by the \Dirac{} program. \\[0.1cm]
     
\procref{Dirac\_Slater\_radial()}
   & Evaluates the Slater integral (\ref{non_rel_Slater_integrals})
     with nonrelativistic orbital functions. \\[0.1cm]

\procref{Dirac\_spin\_orbital()} 
   & Returns the Dirac  spherical spinor $\Omega_{\kappa m}$ from Eq.\
     (\ref{rel_spin_orbitals}). \\[0.1cm]
     
\procref{Dirac\_spinor()}  
   & Returns the relativistic hydrogenic spinor
     $\psi_{n (E) \, \, \kappa m} (r, \vartheta, \varphi)$ from Eq.\
     (\ref{rel_wave_function_spinor}). \\[0.1cm]
     
\procref{Dirac\_Ylm()}       
   & Return the spherical harmonic $Y_{lm}(\vartheta, \varphi)$. \\[0.1cm]
\hline \hline 
\end{tabular}
\end{center}
\end{small}
\end{table}
\subsection{Getting started with D{\small IRAC}} 
\label{organization_started}

The \Dirac{} package is distributed as a compressed tar--file 
{\tt dirac.tar} from which the {\tt dirac} root directory is obtained by 
the command {\tt tar -xvf dirac.tar}. This root contains the source code 
library (for \textsc{Maple} 8), a {\tt Read.me} for the installation of the 
package as well as the program manual {\tt dirac-commands.pdf}. This 
latter document explains all the user relevant commands along with
the output format, their optional arguments and some additional information 
which might be of interest for the application of the procedures. 
The {\tt dirac} root directory also contains an example of a {\tt .mapleinit} 
file which can be easily modified and incorporated into the user's {\tt home}. 
Making use of such a {\tt .mapleinit} file, the module {\tt dirac} should 
then be available like any other module of \textsc{Maple} simply by typing
\begin{flushleft}
{\tt
$ > $ with(dirac); \\
}
\end{flushleft}
\begin{verbatim}      
                            Welcome to Dirac

                      Dirac_save_nuclear_charge := Z 
                 Dirac_save_framework := nonrelativistic 
                       Dirac_save_units := atomic 
\end{verbatim}

As seen from the printout above, the settings of the \Dirac{} program
are specified by the three \textit{global} variables which must not be
overwritten interactively. The (global) variable {\tt Dirac\_save\_framework}, 
for instance, specifies the theoretical \textit{framework} in which the 
various commands will respond to the user; it may take the two values 
{\tt relativistic} and {\tt nonrelativistic} and determines of how
the input and output of the procedures is to be treated internally. In practice,
of course, the input/output data of the procedures depend not only on the
framework but often also on the present choice of the 
\textit{units} as specified by the variable \texttt{Dirac\_save\_units}.
In the \Dirac{} program, we presently support atomic Hartree 
($\hbar$=$m_e$=$e^2 / 4 \pi \epsilon_0$ = 1), natural ($m_e$=$c$=$e$=1), SI, and 
Gaussian units which
are recognized by setting \texttt{Dirac\_save\_units} to one of its
allowed values:  {\tt atomic}, {\tt natural}, {\tt SI}, or {\tt cgs},
respectively. 
Moreover, most of the solutions and output of interest will 
depend on the \textit{nuclear charge} $Z$ as internally specified by 
the global variable {\tt Dirac\_save\_nuclear\_charge}. To change these
particular specifications of the \Dirac{} program, the command 
\procref{Dirac\_settings()} can be used; if instead of 'atomic' units, 
the user wishes to employ 'natural' units for the input and output of the
procedures, the initial specification can be modified by entering 
\begin{flushleft}
{\tt
$ > $ Dirac\_settings(units, natural); \\
}
\end{flushleft}
\begin{verbatim}      
                         Units are changed to natural
\end{verbatim}

Therefore, by specifying the global variables properly due to the demands of the
user, the procedures of the \Dirac{} program can be adapted to the particular 
task. A proper choice of these global definitions makes the program appropriate
for a wide range of applications.

\newpage
\section{Interactive work with the D{\small IRAC} program} 
\label{examples}

To illustrate the use of the \Dirac{} program and to provide some test cases,
below we will display and explain a few examples concerning the fine--structure
of the hydrogen--like ions as well as the (algebraic) access to their
wave functions. A more advanced example later shows the computation of matrix 
elements within the relativistic framework as they arise frequently in the 
study of light--matter interactions. For all these examples, we assume the 
\Dirac{} program to be loaded into the current \textsc{Maple} session as 
explained in the previous section \ref{organization_started}.

\subsection{Fine structure of hydrogen--like ions} 
\label{examples_energy}

In the nonrelativistic Schr\"o{}dinger theory, the level energies 
(\ref{non_rel_energy}) only depend on the principal quantum number $n$ and,
hence, are \textit{degenerate} with respect to the orbital quantum number $l$ 
and the magnetic quantum number $m$. This $2\,n^2$ degeneracy (the factor 2 
arises from the spin of the electrons), however, is partially removed 
if Dirac's theory is applied. In the relativistic case (\ref{rel_energy}),
namely, the nonrelativistic levels with energy $E_n$ split into $n$ different 
sublevels according to the total angular momentum 
$j = 1/2, \, 3/2, ... \, n - 1/2$ of the state. This splitting, which is known 
as the fine--structure of the hydrogen--like ions, can be easily illustrated 
below by means of the \Dirac{} program.

\medskip

To obtain insight into this splitting as a function of the nuclear
charge $Z$, let us first compare the relativistic and nonrelativistic level 
energies for $n = 2$. Since the \textit{nonrelativistic} framework is 
predefined whenever the  \Dirac{} program is (re--) loaded, we start with the
Schr\"odinger energy (\ref{non_rel_energy}) which is returned for an 
arbitrary nuclear charge $Z = Z_0$ by entering
\begin{flushleft}
{\tt
$ > $ Dirac\_settings(charge, Z0); \\
}
\end{flushleft}
\begin{verbatim}      
                        Nuclear charge is changed to Z0
\end{verbatim}
\begin{flushleft}
{\tt
$ > $ E\_2(Z0) := Dirac\_energy(2); \\
}
\end{flushleft}
\begin{verbatim}      
                                     2
                                   Z0
                      E_2(Z0) := - ---
                                    8
\end{verbatim}
Of course, the well--known nonrelativistic energy expression 
$E_{\,n=2} \:=\: - Z_0^{\,2} / 8$ is returned, given in \textit{atomic} units
as the default setting. Analogue, we may obtain the $n \,=\, 2$ energy levels 
as predicted by the Dirac theory. To this end, we first need to change to the
\textit{relativistic} framework for all subsequent calls of the \Dirac{}
procedures
\begin{flushleft}
{\tt
$ > $ Dirac\_settings(framework, relativistic); \\
}
\end{flushleft}
\begin{verbatim}      
                    Framework is changed to relativistic
\end{verbatim}
and, again, make use of the same command {\tt Dirac\_energy()}.
In the relativistic framework, this procedure now expects the two arguments
$n$ and $\kappa$ to return the level energy $E_{n \kappa}$ from Eq.\
(\ref{rel_energy}). To get the energy of a 2$s_{1/2}$ electron with quantum
numbers n=2 and  $\kappa$ = -1, we type
\begin{flushleft}
{\tt
$ > $ E\_2s(Z0) := Dirac\_energy(2, -1); \\
}
\end{flushleft}
\begin{verbatim}      
                                          1
                  -------------------------------------------------- - 1
                  /              2                     2        \1/2
                  |            Z0  Dirac_Constant_alpha         |
                  |1 + -----------------------------------------|
                  |                                  2   2 1/2 2|
                  \    (1 + (1 - Dirac_Constant_alpha  Z0 )   ) /
      E_2s(Z0) := ------------------------------------------------------
                                                      2
                                  Dirac_Constant_alpha
\end{verbatim}
which returns this energy in the most general form, including the nuclear
charge Z0 and the (unevaluated) fine structure constant $\alpha$ as yet given 
by the global variable \newline {\tt Dirac\_Constant\_alpha}.
At the first glance, such a general form may look quite complicated but it 
helps of course in treating the expressions algebraically. For our discussion 
of the fine--structure behaviour as a function of $Z$, we may first assume the 
$\alpha Z$ parameter small when compared with unity and apply the standard 
\textsc{Maple} command {\tt series()} to expand this expression for the
2$s_{1/2}$ energy in powers of $Z$ 
\begin{flushleft}
{\tt
$ > $ series(E\_2s(Z0), Z0): \\
$ > $ E\_2s\_exp(Z0) := convert(\%, polynom);
}
\end{flushleft}
\begin{verbatim}
                              2                             2   4
        E_2s_exp(Z0) = -1/8 Z0  - 5/128 Dirac_Constant_alpha  Z0
\end{verbatim}
Subtracting from this expression the Schr\"odinger energy $E_{\,n=2}$ from 
above
\begin{flushleft}
{\tt
$ > $ E\_2s\_exp(Z0) - E\_2(Z0);
}
\end{flushleft}
\begin{verbatim}
                                             2   4
                       5 Dirac_Constant_alpha  Z0
                     - ---------------------------
                                   128
\end{verbatim}
we find the so--called relativistic Pauli correction $\Delta E_{\,n \kappa}$ to
the (nonrelativistic) level energy \cite{BeS57, BaJ83} which, as seen from the
printout, scales like $\alpha^2 \,Z_0^{\,4}$ and, thus, will strongly increase
for high--Z ions.   

\medskip

The result above for the Pauli correction $\Delta E_{n \kappa}$ displays the
analytic behaviour of the fine--structure splitting; to get a better impression
on the size of this splitting, we may evaluate this expression for $n \,=\, 2$
also \textit{numerically}. To substitute the value $\alpha$ = 1/137.036 of the
fine--structure constant for {\tt Dirac\_Constant\_alpha}, we may use the
command {\tt Dirac\_evalf()} in the \Dirac{} program and obtain
\begin{flushleft}
{\tt
$ > $ E\_2s(Z0) := Dirac\_evalf(E\_2s(Z0)); \\
}
\end{flushleft}
\begin{verbatim}
                                    18778.865231694104538
  E_2s(Z0) := ---------------------------------------------------------- 
             /                                           2         \1/2
             |              0.000053251353990881516089 Z0          |
             |1. + ------------------------------------------------|
             |                                              2 1/2 2|
             \     (1. + (1. - 0.000053251353990881516089 Z0 )   ) /
                                                                                
                                                                                
            - 18778.865231694104538
\end{verbatim}

By evaluating in a similar way also the level energy for the 2p$_{3/2}$ shell
($n$ = 2, $\kappa$ = -2) 
\begin{flushleft}
{\tt
$ > $ E\_2p(Z0) := Dirac\_energy(2, -2); \\
$ > $ E\_2p(Z0) := Dirac\_evalf(E\_2p(Z0)): \\
}
\end{flushleft}
we can finally calculate the level shifts of the $j = 1/2$ and $j = 3/2$
fine--structure levels with respect to the value from the nonrelativistic 
theory. For atomic hydrogen ($Z_0$ = 1), this gives rise to
\begin{flushleft}
{\tt
$ > $ subs(Z0 = 1, E\_2s(Z0) - E\_2(Z0)); \\
$ > $ subs(Z0 = 1, E\_2p(Z0) - E\_2(Z0)); \\
}
\end{flushleft}
\begin{verbatim}
                                             -5
                        -0.208018917100000 10
       
                                            -5
                        -0.41602897100000 10
\end{verbatim}
in atomic units and can be compared immediately to the values 
of 0.46 cm$^{-1}$ (0.208 $\times$ 10$^{-5}$ au) and 
0.09 cm$^{-1}$ (0.416 $\times$ 10$^{-6}$ au), respectively, as known from the
literature \cite{BeS57, BaJ83}.

\subsection{Relativistic contraction of the s--orbitals} 
\label{examples_wavefunctions}

In section 2, we have displayed the radial components for a bound--state
electron for both, the nonrelativistic (\ref{non_rel_wave_function_bound})
and relativistic theory (\ref{hydrogenic-Pnkappa}--\ref{hydrogenic-Qnkappa}).
The spatial behaviour of these one--electron solutions for different 
nuclear charges $Z$ has been discussed in detail at a number of places
\cite{BeS57, EiM95} to illustrate the 'relativistic effects' for high--$Z$ 
ions. While, for small charges $Z$, the radial Schr\"odinger function  
(\ref{non_rel_wave_function_bound}) and 
the large component (\ref{hydrogenic-Pnkappa}) from the Dirac theory are 
practically the same, a comparison of these functions for high--Z ions shows
a clear contraction of the large component towards the 
nucleus. This effects, which is known as the \textit{relativistic contraction 
of the wavefunctions}, influences not only the structure of heavy 
(hydrogen--like) ions but, for instance, also their ionization behaviour
\cite{KoF03a} or the chemical binding in materials containing heavy elements.

\medskip

To illustrate the contraction of the $3s$ wavefunction for a hydrogen--like 
uranium ion $U^{\,91+}$ by means of the \Dirac{} program, we shall extract and
compare the radial function (\ref{non_rel_wave_function_bound}) with the large
component (\ref{hydrogenic-Pnkappa}) from Dirac's theory. The algebraic 
expression for the $P_{\,n = 3, \, l = 0}\,(r)$ nonrelativistic function can 
be obtained by 
\begin{flushleft}
{\tt
$ > $ Dirac\_settings(framework, nonrelativistic); \\
}
\end{flushleft}
\begin{verbatim}      
                Framework is changed to nonrelativistic
\end{verbatim}
\begin{flushleft}
{\tt
$ > $ Dirac\_settings(charge, 92); \\
}
\end{flushleft}
\begin{verbatim}      
                    Nuclear charge is changed to 92
\end{verbatim}
\begin{flushleft}
{\tt
$ > $ P\_3s(r) := Dirac\_radial\_orbital(3, 0, r); \\
}
\end{flushleft}
\begin{verbatim}      
                 368   1/2  1/2       92 r                        2
      P_3s(r) := --- 23    3    exp(- ----) (27 - 1656 r + 16928 r ) r
                 243                   3
\end{verbatim}
where, in the first two lines of printout, we set the nonrelativistic framework
and define the nuclear charge to be $Z$ = 92.

\medskip

Returning to the relativistic framework as before, the large component
of a $3s_{1/2}$ electron is generated in the \Dirac{} program by calling 
\begin{flushleft}
{\tt
$ > $ P\_3s\_r(r) := Dirac\_radial\_component({\it large}, 3, -1, r): \\
}
\end{flushleft}
for which, we suppress here the printout to screen (using the colon 
at the end of the line) as this would result in quite a lengthy expression in
terms of the hypergeometric function and various physical constants 
[cf.\ Eq.\ (\ref{hydrogenic-Pnkappa})]. For the present purpose, instead, 
we first better 'evaluate' all the physical constants to their numerical 
values by calling
\begin{flushleft}
{\tt
$ > $ P\_3s\_r(r) := Dirac\_evalf(P\_rel(r)); \\
}
\end{flushleft}
\begin{verbatim}      
                                      0.74113463002424792093
  P_3s_r(r) := 69.235282776899202095 r

      exp(-32.599243267231110742 r) (3.8221513992160292467

      hypergeom([-2.], [2.4822692600484958419], 65.198486534462221484 r) -

      2. hypergeom([-1.], [2.4822692600484958419], 65.198486534462221484 r))
\end{verbatim}
Having the $3s$ radial wave functions from the nonrelativistic and relativistic
theory, we can plot them by using the standard \textsc{Maple} procedures 
{\tt plot()}. Physically more important than the wave function, however, 
is the 'contraction' in the charge distribution as given by the 
\textit{radial distribution functions} $|P_{n=3, \: l=0}(r)|^2$ and 
$|P_{n=3, \: \kappa = -1}(r)|^2$, respectively.
\begin{flushleft}
{\tt
$ > $ plot([(P\_3s(r))\^{}2, abs(P\_3s\_r(r))\^{}2], r=0..0.3, linestyle=[DASH, SOLID]);
}
\end{flushleft}
\begin{center}
\epsfig{file=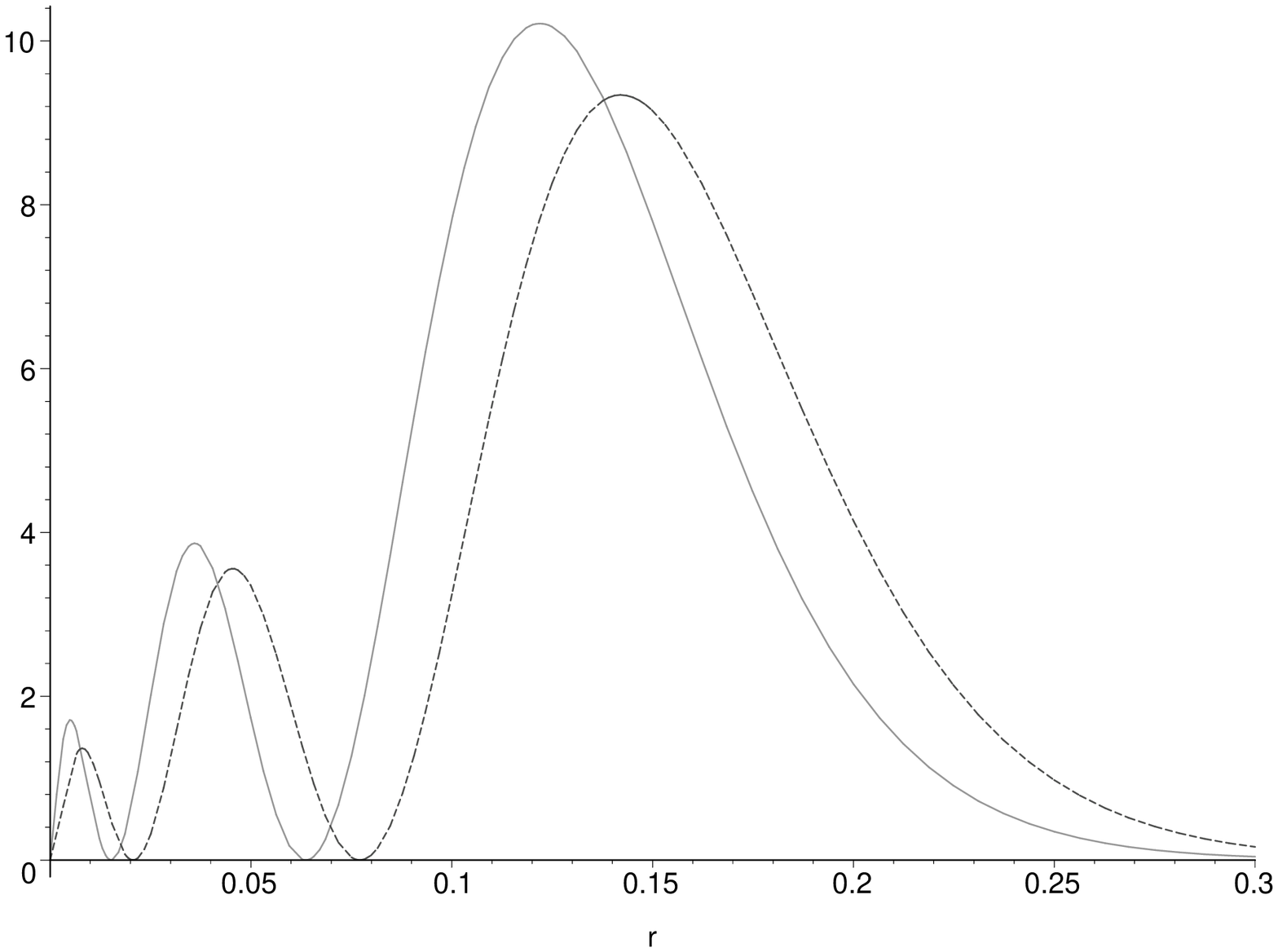, height=10cm, angle=270}
\end{center}
As expected, a clear contraction of the relativistic function (solid line)
is found when compared with the nonrelativistic radial wavefunction
(dashed line).

\subsection{Evaluation of the relativistic radial integrals} 
\label{examples_radiative}

Until now, we have demonstrated the use of the \Dirac{} program by simple
'textbook examples' such as the computation of the level energies or the
hydrogenic wave functions. The \Dirac{} procedures, however, are able to help
the user with much advanced studies concerning the excitation and decay
properties or the dynamical behaviour of the hydrogen--like ions. For such
investigations, we need to compute \textit{transition matrix elements} including
the wavefunctions from above and the corresponding (transition) operators. 
The exact relativistic treatment of the \textit{photoionization} of 
hydrogen--like ions, for instance, can be traced back to the computation
of the bound--free transition amplitude \cite{EiM95}
\begin{equation}
   \label{photoionization}
   M^{\rm\, PI}_{f, b} \;=\; 
   \mem{\psi_{\,E_f \kappa_f m_f}}{ \bm{\alpha} \mathbf{ \, u}_{\lambda} 
        e^{i \bm{k}\bm{r}}}{\psi_{\,n_b \kappa_b \mu_b}} \, ,
\end{equation}
where $\psi_{\,n_b \kappa_b \mu_b} (\mathbf{r})$ and 
$\psi_{\,E_f \kappa_f m_f} (\mathbf{r})$ are the (relativistic) wavefunctions 
for the initially bound and the final continuum electron, respectively,
and where the operator ${\mathbf u}_{\lambda} e^{i \bm{kr}}$ 
describes the electron--photon interaction.

\medskip

The simplification of the transition amplitude (\ref{photoionization})
requires elements from the algebra of angular momentum and has been discussed 
elsewhere \cite{EiM95, SuF02}. In particular, it was shown that this amplitude 
can be factorized into radial and angular parts where the latter ones
evaluate to Wigner $3-j$ and $6-j$ symbols and a constant factor,
while the radial part is reduced to the integrals 
(\ref{ILplusminus}--\ref{JL}). Below, we therefore discuss how the \Dirac{}
package can be utilized to compute these radial integrals. We start with the
straightforward numerical integration of the radial integral
$\,I_{\,L = 1}^{\,0} (q;a,b)\,$, which arises if a 1s$_{1/2}$ electron 
($n_b = 1$, $\kappa_b$ = --1) is ionized by a photon with energy 
$E_\gamma \,=\, 70$ eV into a $s_{1/2}$ continuum state (partial $s-$wave; 
$\kappa_f$ = --1). For this particular choice of the parameters and quantum 
numbers, we may simply enter
\begin{flushleft}
{\tt
$>$ Dirac\_settings(charge, 1): \\
$>$ Dirac\_settings(framework, relativistic): \\
$>$ Pa(r) := Dirac\_evalf(Dirac\_radial\_component({\it large}, 1, -1, r)): \\
$>$ Qb(r) := Dirac\_evalf(Dirac\_radial\_component({\it small}, 
                          2.072432345, -1, r,{\it free})): \\
$>$ int(Pa(r)*Dirac\_spherical\_j(1,0.01877198124*r)*Qb(r),r=0..70.); \\
}
\end{flushleft}
\begin{verbatim}      
                                                  -5
                        -0.23633058961153019500 10
\end{verbatim}
where the keyword \textit{free} in the procedure 
\procref{Dirac\_radial\_component()} is used to generate the small component of
a continuum state with (kinetic) energy $E_f$ = 2.072432345 (au) and where most of the 
intermediate printout has been 
suppressed by using a colon to terminate the first 5 lines. 
Both the energy $E_f$ and the momentum transfer
$q$ = 0.01877198124 (au) are easily derived from the conservation of energy,
assuming $I$ = 0.5 (au) as the ionization potential for the $1s$ electron of
atomic hydrogen. In the last input line, moreover, the procedure 
\procref{Dirac\_spherical\_j()} is used to returns the spherical
Bessel function $j_L (qr)$.

\medskip

The (direct) numerical computation of the integrals 
(\ref{ILplusminus}--\ref{JL}) is typically reliable but also very 
\textit{time consuming}. For the integral above, for instance, \textsc{Maple}
requires about 145 seconds at a 1 GHz processor. An alternative and (much)
faster computation is achieved by using the \textit{analytical} 
expressions for the radial integrals (\ref{ILplusminus}--\ref{JL}) in terms of
the special hypergeometric functions \cite{TrB83}. Within the \Dirac{} program,
such a (analytical) integration algorithm is supported by means of the 
procedure \procref{Dirac\_IJ\_radial()} which can be invoked with the same
parameters as above to determine the $\,I_{\,L = 1}^{\,0} (q;a,b)\,$ integral
\begin{flushleft}
{\tt
$>$ Dirac\_IJ\_radial(IL0, 0.01877198124, 1, 1, -1, 2.072432345, -1, free): \\
}
\end{flushleft}
\begin{verbatim}      
                                                  -5
                        -0.23633058961153083452 10                                                  
\end{verbatim}
Although, of course, the integral is the same for using both, the numerical
or analytical integration scheme (at least up to the 16--th decimal digit),
the procedure \procref{Dirac\_IJ\_radial()} is faster by almost two orders of
magnitude when compared to \textsc{Maple}'s 
numerical integration. It is also more
accurate as found by increasing the
number of digits to \texttt{Digits := 30}.

\section{Summary and outlook} 
\label{outlook}

The \Dirac{} program has been presented to facilitate a fast and consistent
access to the basic elements and properties of the hydrogen--like ions. 
In its present form, this program consists out of a set of \textsc{Maple} 
procedures for the numerical and analytical evaluation of the Coulomb wave 
and Green's functions, both within the nonrelativistic and relativistic 
framework. Apart from these functions, moreover, the computation of radial 
integrals (as they frequently arise in the evaluation of transition amplitudes
for various types operators) is supported by means of the \Dirac{} program by 
implementing the analytic formulas.

\medskip

Of course, the procedures above can be seen also as the \textit{technical basis}
for more advanced studies concerning the behaviour of the hydrogen--like 
ions or, more generally, applications of the atomic shell model. In the future,
therefore, several lines are possible (and desirable) for the further
development of the \Dirac{} program. In a first step, we want to extent the
present set of procedures towards the interaction of hydrogen--like ions with 
the radiation filed as it occurs in the ionization or capture of electrons
by high-$Z$ ions or in laser fields. Examples of very desirable procedures
refer to a simple access to the decay rates, angular distributions, or to
the polarization of the emitted radiation. Having collaborations with
the atomic physics community at GSI and elsewhere for many years, we know 
that these developments will make the \Dirac{} package attractive to both, 
experimental and theoretical investigations on energetic ion--atom and 
ion--electron collisions. 

\section*{Acknowledgments}
    
We gratefully acknowledge Thorsten Inghoff for valuable suggestions
and help with the development of the code.         
This work has been supported by the BMBF in the framework of the 
\textit{'Verbundforschung im Bereich der Hadronen-- und Kernphysik'}.


\end{document}